%% file: main.tex
\documentclass[11pt]{article}

\usepackage[T1]{fontenc}
\usepackage[margin=1in]{geometry}
\usepackage{lmodern}
\usepackage{amsmath}
\usepackage{amssymb}
\usepackage{graphicx}
\usepackage{tikz}
\usetikzlibrary{calc,positioning,arrows.meta}
\usepackage{booktabs}
\usepackage{tabularx}
\usepackage{caption}
\usepackage{microtype}
\usepackage{xcolor}
\usepackage{xspace}
\usepackage[numbers,sort&compress]{natbib}
\usepackage[hidelinks]{hyperref}
\hypersetup{
  pdftitle={Pretraining and adapting a language model on a dependency-free stack},
  pdfauthor={Thang Tran, Lan Dang},
}

\graphicspath{{figures/}}
\input{results}

\newcommand{\numbat}{\emph{numbat}\xspace}

\title{{\huge\bfseries Pretraining and adapting a language model on a dependency-free stack}\\[8pt]
{\large\normalfont GPT-2 124M from random weights, reproduced against llm.c,\\
and a clinical adapter for Qwen3-0.6B}}
\author{%
  Thang Tran\thanks{Corresponding author: \texttt{thang.tran@cloudkites.com}}\\
  \small CloudKites AI Lab\\
  \small New South Wales, Australia\\
  \small\texttt{thang.tran@cloudkites.com}
  \and
  Lan Dang\\
  \small Monash Business School, Monash University\\
  \small Victoria, Australia\\
  \small\texttt{LanHong.Dang@monash.edu}
}
\date{September 2026}

\begin{document}
\maketitle

\begin{abstract}
Almost every language model in service was trained by one family of software. That
concentration makes a question hard to settle: how much of what is known about
training a language model describes language models, and how much describes that
software? Settling it needs a second implementation able to carry a model through a
whole lifecycle rather than reproduce one operator.

We report such a lifecycle. Using \numbat{}, a machine-learning stack written in Zig
with no third-party runtime dependencies, we pretrain a \NParamsM{}-parameter GPT-2
from random initialisation over \TokensTotal{} tokens of web text, then adapt a
separate small model to clinical question answering. A reference implementation runs
on identical hardware at both stages, and a sidecar with authority to halt a run
supervises each.

Agreement is close. Held-out cross-entropy finishes at \LastValLoss{} against a
published \TargetValLoss{}, and HellaSwag at \LastHella{} against \TargetHella{};
across \NPaired{} paired evaluations it sits below a same-machine reference at every
point, by \MeanDevAbs{} on average. Re-running that reference on different hardware
moves it \RefCrossDelta{}, which bounds how much of any gap is method rather than
framework. Throughput does not pay for agreement: measured in one session at a
production configuration, \numbat{} reaches \SdkTps{} tokens per second against
PyTorch's \SdkPtTps{}, scaling \SdkSpeedup{} over three cards. Clinical adaptation
ends at \ClinNbLoss{} held-out loss against \ClinPtLoss{}.

Neither model is a medical device, and neither is validated for clinical use.
\end{abstract}

\section{Introduction}\label{sec:intro}

Training a language model is a long chain of arithmetic. Tokens are read from disk,
embedded, pushed through a stack of attention and feed-forward blocks, scored against a
target, differentiated, reduced across devices, and folded into a weight update. Then it
happens again, tens of thousands of times. A mistake anywhere in that chain rarely announces itself.
Loss still falls. Generated text still reads fluently. What changes is how good the model
finally becomes, and by how much is not knowable from inside the run.

This is an ordinary testing problem with an awkward property: for most of the computation
there is no independently known right answer. Software engineering's usual answer is to build
a second implementation and compare. For language-model training, though, that answer has
been largely theoretical, because a second implementation complete enough to pretrain a model
end to end is a large thing to build, and almost all of this work runs on one family of
Python-orchestrated frameworks.

We have been building such an implementation and reporting on it as it grew. An earlier study
used a clinical fine-tune to test whether two independent stacks agree step by step, and
catalogued seventeen faults that comparison exposed~\citep{tran2026crossstack}. A companion
paper described construction and verification of the stack itself, using a vision detector
trained from scratch on COCO as its acceptance test~\citep{tran2026numbat}. A third applied it
to endoscopy, training a foundation model without labels and fine-tuning task models from
it~\citep{tran2026woma}.

One thing none of those did is pretrain a language model. That gap matters, because
pretraining is where a training stack is hardest to get right and easiest to be quietly wrong
about. It runs for days rather than minutes. It uses mixed precision, gradient accumulation
and multi-device reduction all at once. Its errors compound over tens of thousands of updates.
Unlike a fine-tune, pretraining starts from noise, so no pretrained checkpoint carries
the model most of the way regardless of what the software does.

\paragraph{What we did.}
We reproduce a published pretraining run---llm.c's GPT-2 at \NParamsM{}
parameters~\citep{karpathy2024llmc,radford2019gpt2}---on \numbat{}, matching architecture,
corpus, tokenizer, precision, optimizer and schedule, and differing in one place we state
rather than hide (\S\ref{sec:geometry}). We then carry the same stack into a medical domain,
adapting a small instruction model to clinical question answering, because a framework that
trains a general model well and cannot be trusted on a specialist one has not been shown to
be much.

\paragraph{Contributions.}
\begin{enumerate}
\item \textbf{A language-model pretraining run on an independent stack, checked against a
reference on the same machine} (\S\ref{sec:method}, \S\ref{sec:results}). We describe how a
reference trajectory is produced, what it costs, and what it can and cannot settle. Over
\NPaired{} paired evaluations held-out loss sits below that reference at every point.
\item \textbf{Evidence about how far a reference trajectory is itself reproducible}
(\S\ref{sec:refcross}). We ran a second reference on different hardware, with a different
number of cards and a slightly different batch, and it agrees with the first to
\RefCrossDelta{} at step \RefCrossStep{}. Without this control, any framework gap and any
artefact of how the reference was run are the same number.
\item \textbf{A control that can act, not only observe} (\S\ref{sec:control}). Streaming
metrics to a dashboard does not stop a run that has gone wrong; a person noticing does, and
people sleep. We describe a sidecar with authority to halt, the arithmetic that keeps its
judgement honest once a reference runs out, and its record across both stages.
\item \textbf{A speed account at the configuration a real run uses} (\S\ref{sec:speed}), with
each remaining gap attributed to a named mechanism rather than left as a total.
\end{enumerate}

\paragraph{What we do not claim.} We do not claim a new model, a new training method, or a
better model than the reference. Reproduction is the point: a result already known is what
makes a second implementation checkable. On the clinical side we claim nothing about answer
quality. Lower held-out loss on clinical text is not evidence that a model answers clinical
questions correctly, and \S\ref{sec:limits} says so at greater length. The adapted model is a
research artefact.

\section{Background}\label{sec:background}

\paragraph{The model and its reproduction.} GPT-2 is a decoder-only
transformer~\citep{vaswani2017attention,radford2019gpt2}. Its smallest configuration, at
around 124 million parameters, has become a common reference point for training work because
it is small enough to train on a modest budget and well enough documented to check. llm.c is a
reproduction of that training in C and CUDA, together with a PyTorch trainer that implements
the same recipe~\citep{karpathy2024llmc}. Its reported result for the smallest configuration,
after \LlmcTokensTotal{} tokens of FineWeb, is a held-out loss near \TargetValLoss{} and a
HellaSwag score near \TargetHella{}. We take that as our target.

Two properties make llm.c a good reference rather than a convenient one. Its recipe is
written down completely, including choices usually left implicit: which tensors receive
weight decay, how the vocabulary is padded, what the softmax covers. It also ships a reference
dump: weights, activations, gradients and a short optimizer trace from a fixed batch, against
which another implementation can be checked before it trains anything at all.

\paragraph{Corpus.} FineWeb is a filtered and deduplicated corpus of web
text~\citep{penedo2024fineweb}. We use its ten-billion-token sample, tokenized with GPT-2's
byte-level encoding~\citep{sennrich2016bpe}. That budget is roughly eighty tokens per
parameter, well past the ratio near twenty that~\citet{hoffmann2022chinchilla} derive as
compute-optimal, and in the regime~\citet{kaplan2020scaling} describe where further data
keeps buying loss at a diminishing rate. We inherit it from llm.c rather than choosing it:
matching a published run means matching its budget.

\paragraph{Measurement.} Two numbers are reported throughout. Held-out cross-entropy is
measured on a fixed slice of the corpus never trained on. HellaSwag is a multiple-choice
benchmark of sentence completion~\citep{zellers2019hellaswag}; we score it in the completion
style llm.c uses, where each candidate ending is scored by length-normalised likelihood.

\paragraph{Method.} Optimisation is AdamW~\citep{loshchilov2019adamw} with a cosine
schedule~\citep{loshchilov2017sgdr}, mixed precision in the arrangement introduced
by~\citet{micikevicius2018mixed}, and data-parallel training across
devices~\citep{li2020pytorchddp}. The clinical stage uses low-rank
adaptation~\citep{hu2021lora} with rank-stabilised scaling~\citep{kalajdzievski2023rslora}.

\section{The stack under test}\label{sec:stack}

\numbat{} is a machine-learning stack written in Zig~\citep{zig}, described at length
in~\citet{tran2026numbat}. Four of its properties bear on this study.

\textit{It is self-contained.} Tensors, automatic differentiation, layers, mixed precision,
data loading, multi-device reduction, checkpointing and monitoring are one library with no
third-party runtime dependency. Nothing in the chain between disk and weight update comes from
the stack being compared against, so agreement is evidence rather than shared provenance.

\textit{A run is a declarative artefact.} Architecture, optimisation, batch geometry,
evaluation cadence and output paths live in one recipe file, which is validated before any
work starts and recorded with results. This matters for reproduction: the recipe in
\S\ref{sec:method} is the run, not a description of it.

\textit{It has two front doors onto one core.} A model can be trained through the framework's own
trainer, which imports the library directly, or through a C ABI that a binding in any language
composes from---the path our SDK takes. Both drive the same tensors, kernels, autograd and
collectives. That is what makes the comparison in \S\ref{sec:speed} informative in both
directions: the two doors should agree, and a kernel-level profile of them at identical
geometry finds the same kernels in the same counts, so a measured difference between them is
evidence about measurement rather than about the stack. It is also what produced the
retraction reported there. Until the C ABI exposed the recipe plane, the two doors read
\emph{different} declarations of the same run---the framework trainer a validated recipe file,
the SDK constants compiled into it---and a divergence that had gone unnoticed for months
appeared in the one place a declaration would have made it visible before the first step.

\textit{It is built to be driven by an agent as well as a person.} The command-line surface
reports what a build can do, raises typed diagnostics, and returns exit codes that distinguish
a failed measurement from a malformed request; acceptance gates are commands with verdicts and
the number that decided them. We mention this because it shapes how the run below is
controlled---checks are commands anything can run and act on, rather than plots for a human
to read.

\section{Method}\label{sec:method}

\subsection{Model}\label{sec:model}

\input{fig_architecture}

Figure~\ref{fig:architecture} shows what is trained. We take GPT-2's smallest
configuration exactly: \NLayer{} transformer blocks, \NHead{}
attention heads, \NEmbd{} embedding channels, context \BlockSize{} tokens, and input and
output embeddings tied. Each block applies layer normalisation before attention and before
its feed-forward part, with residual connections around both. Parameters total
\NParams{}.

Two details are easy to get wrong and both were taken from llm.c rather than assumed.
Vocabulary is padded from \VocabReal{} entries to \VocabPadded{}, a multiple of 128, because
matrix shapes that are multiples of 128 select faster kernels; but the softmax covers only the
\VocabReal{} real tokens, so padding rows receive no probability and no gradient. Weights are
initialised from a normal distribution with standard deviation 0.02, and projections that
write into a residual stream are scaled down by $1/\sqrt{2L}$ for $L$ blocks, which keeps
variance of that stream stable as depth grows.

\subsection{Corpus and tokenisation}\label{sec:data}

\input{fig_pipeline}

Figure~\ref{fig:pipeline} shows the corpus, how it is split, and what each part is
used for. Training data is FineWeb's ten-billion-token sample. Documents are tokenized with GPT-2's
byte-level encoding, each preceded by an end-of-text marker, concatenated in dataset order and
cut into shards of exactly \ShardTokens{} tokens, with a document that crosses a boundary
continuing into the next shard. Shard zero is held out for evaluation; training reads the rest
in order. Held-out data is therefore a contiguous prefix of the corpus and is never trained
on.

\numbat{} implements this preparation itself, and we checked that implementation rather than
trusting it. Running llm.c's preparation script and \numbat{}'s over identical documents and
comparing the first shard byte for byte, token payloads are identical across all
\ShardBytes{} bytes; headers differ in one field, where \numbat{} records vocabulary size and
llm.c writes zero. Shards used by the run itself are llm.c's own, so this comparison
establishes that preparation would have produced the same ones rather than forming part of the
run.

\subsection{Optimisation and precision}\label{sec:optim}

Optimisation follows llm.c without deviation: AdamW with $\beta$ of \AdamBetas{}, epsilon
$10^{-8}$, weight decay \WeightDecay{} applied to two-dimensional weights only, and gradients
clipped to global norm \GradClip{}. Learning rate rises linearly over \WarmupSteps{} steps to
\PeakLR{} and decays by a cosine curve to zero across all \TotalSteps{} steps. Random seed is
\Seed{}.

Precision follows the same source. Activations are bfloat16; master weights, gradients and
optimizer state stay in 32-bit floating point; normalisation statistics and loss are computed
in 32-bit. This is the arrangement of~\citet{micikevicius2018mixed} without loss scaling,
which bfloat16's exponent range makes unnecessary.

Gradient spike skipping is switched off, because llm.c skips no steps. Stated plainly: a stability feature left enabled would have made our run more robust than the run we
claim to reproduce.

\subsection{Batch geometry, and the one place we differ}\label{sec:geometry}

llm.c consumes \LlmcTokensPerStep{} tokens per optimizer step on eight cards. Our machine has
three. Because \LlmcTokensPerStep{} divided by three cards and \SeqLen{} tokens is not a whole
number, no three-card arrangement reaches that batch exactly. We use \WorldSize{} cards,
\PerDeviceBatch{} sequences each, \GradAccum{} accumulation steps: \TokensPerStep{} tokens per
step, \BatchDiffPct{} larger.

Step count, warmup and schedule are llm.c's, so our run consumes \TokensTotal{} tokens rather
than \LlmcTokensTotal{}, and every curve is compared by step. We state this rather than
rounding it away because it is the only respect in which our recipe is not llm.c's, and
because its direction is known: slightly more data per step, which if anything flatters us.

Evaluation is matched the same way. llm.c scores 20 batches of $64\times1024$ on each of eight
cards every \ValEvery{} steps, averaged across cards: \LlmcValTokens{} tokens. We score 180
batches of $19\times1024$ on each of three, averaged the same way: \ValTokens{} tokens, larger
in the same proportion as the batch.

\subsection{What is checked before training starts}\label{sec:pregates}

A reference trajectory tells you that a run has gone wrong, eventually. Checks that run in
seconds tell you before you spend days. Three run first.

\textit{Against llm.c's reference dump, at four levels.} Loading llm.c's weights and its fixed
batch, we compare forward activations, loss, every one of the \NGradTensors{} parameter
gradients, and ten AdamW steps. All four are compared because each catches something the
others miss: matching loss with a wrong backward is possible, and so is matching both with a
wrong optimizer.

\textit{Corpus preparation}, as described in \S\ref{sec:data}.

\textit{The scorer, against a known model.} A benchmark implemented slightly differently
scores differently, and that difference would then be attributed to training. Before training
anything we score OpenAI's released GPT-2 with our own implementation and compare against
llm.c's script on the same checkpoint and the same \HellaItems{} items.

\subsection{The reference trajectory}\label{sec:reference}

A long run with no reference and no acting gate is a smoke test that happens to be expensive.
Our reference is llm.c's own PyTorch trainer, run on the same three cards, with the same
corpus, schedule and precision.

We do not copy llm.c's file. It is read, three changes are applied in memory---each asserted
to match exactly once, so that an upstream edit fails loudly rather than silently changing
what is being compared---and the result is executed. All three align that PyTorch script with
llm.c's C trainer, which is the run whose published numbers are the target: vocabulary padded
to \VocabPadded{}, softmax over the \VocabReal{} real tokens, and held-out loss averaged over
all cards rather than reported from one.

Reference and run cannot occupy the same cards at the same time. Ours ran first, to step
\RefLastMeasured{} of \TotalSteps{}, and was then stopped so that the cards could carry the
run it gates. This is the central limitation of a same-machine reference on a small budget,
and \S\ref{sec:refcross} describes what we did about it.

\subsection{Control that can act}\label{sec:control}

Metrics on a dashboard are an observation. Stopping a run that has gone wrong is a decision,
and on a multi-day run it is one nobody is awake for. A sidecar process therefore holds that
authority. Every few minutes it reads metrics from disk, compares held-out loss against the
reference at the same step, and on three consecutive evaluations more than \GateBandPct{}
above that reference writes a halt marker and terminates the run's whole process group. Three
consecutive evaluations rather than one, because a single evaluation is noisy; a marker on
disk as well as a signal, because a supervisor that restarts crashed runs must be able to tell
a crash from a deliberate stop.

Once a run passes the reference's last measured step, a band against that reference stops
meaning anything: a curve still falling clears a fixed final value forever. Two mechanisms
cover that stretch honestly.

\textit{Points past the measurement are marked, and marked points only warn.} We fill the
remaining range by interpolating between the last measured point and llm.c's published
endpoint, in logarithm of step count, because loss falls roughly linearly against that and
markedly not against step count. Every such point carries a flag, and a flagged point can
raise an alert but never stop a run. A band whose shape is modelled rather than measured can
be wrong in the tightening direction, and cost of being wrong there is a multi-day run killed
for nothing.

\textit{Authority to stop passes to a check that compares a run only against itself.} If no
evaluation improves on the best seen for a set number of evaluations, that is a stall, and a
stall is judged without reference to anything external.

The gate also refuses comparisons it cannot make. If reference and run scored different
numbers of examples, it says so instead of differencing them---a 512-example evaluation and a
3,440-example evaluation are different measurements, and quietly subtracting one from the
other is how a tool reports a gap that is an artefact of its own inputs.

Around both sits a supervisor that relaunches from the most recent checkpoint after a crash
and stops for good on either a completion marker or a halt marker. Checkpoints are written
every 500 steps with optimizer state, so a restart resumes rather than repeats.

\subsection{The clinical stage}\label{sec:clinical-method}

Pretraining produces a general model. Most useful models are specialists, and specialising one
exercises different machinery: adapter parameters beside frozen weights, a different data
pipeline, masking so that loss is taken on answers rather than questions. We therefore ran a
second stage in a domain where being quietly wrong matters more than usual.

Base model is Qwen3 at 0.6 billion parameters~\citep{qwen3}, adapted with low-rank adaptation
of rank \ClinRank{} and scaling \ClinAlpha{} on seven projections per layer, rank-stabilised,
giving \ClinTrainable{} trainable parameters. Training data is three public question-answering
sets---PubMedQA~\citep{jin2019pubmedqa}, MedMCQA~\citep{pal2022medmcqa} and
MedQA~\citep{jin2020medqa}---whose licences were checked and recorded before use. Conversion
accounts for every record: \ClinRecordsIn{} in, \ClinAccounted{} accounted as converted or
duplicate, with a non-zero exit if those totals disagree. Split is \ClinTrain{} training and
\ClinEval{} held-out examples.

Reference arrangement differs from pretraining, and the difference is
deliberate. Both stacks ran the same recipe on the same three cards, but \emph{one leg
at a time} and each alone on the box: PyTorch with Transformers and
PEFT~\citep{paszke2019pytorch,wolf2020transformers,peft2022} first, then \numbat{},
gated against the curve the first leg had written. Running them concurrently---as
the earlier study did, one card each---buys a shared clock and costs a clean wall
clock, because two trainers on one machine contend for its CPU, its PCIe and its power
budget. Separating them costs the shared clock and buys a second leg that can be gated
against a real reference curve rather than against "loss goes down", which is the
control the pretraining run of \S\ref{sec:res-trajectory} could only have for part of
its schedule. Nothing had to be interpolated because nothing was truncated. The gate
polled every \ClinGatePoll{} minute and had authority to stop the run.

This workload is the one used in~\citet{tran2026crossstack}, and we are careful about what is
new here. That study's contribution was a protocol and a fault inventory. What we report in
\S\ref{sec:clinical} is a later full-epoch run with a same-machine reference leg, used here as
second half of a lifecycle rather than as a fresh claim about differential testing.

\section{Results}\label{sec:results}

Apparatus is three machines, identified throughout by hardware rather than by name.
\textbf{Box~1} is three RTX 3090 cards on one host, capped at \PowerCap{}~W and connected by
PCIe with no direct card-to-card link; it carries the pretraining run and the clinical
adaptation. \textbf{Box~2} is two uncapped cards of the same model, carrying per-stage timing
and the second reference---uncapped, so a per-card figure from it is not throttle-limited.
\textbf{Box~3} is a single RTX 3060, used for the small-batch row where the step is short
enough that launch overhead is the whole of it. The same labels appear in
Table~\ref{tab:endtoend}, and the full description is in \texttt{data/apparatus.json}.

\subsection{Before training: agreement with the reference dump}\label{sec:res-pregates}

Loading llm.c's \CheckParams{} parameters and its fixed batch, largest disagreement in forward
activations is \CheckLogit{}, in loss \CheckLoss{}, and across all \NGradTensors{} parameter
gradients \CheckGrad{}, worst at \CheckWorstTensor{}. After ten AdamW steps, losses differ by
\CheckSteps{}. All four sit inside llm.c's own tolerance of \CheckTol{}.

Corpus preparation is identical over \ShardBytes{} bytes of tokens
(\ShardDocs{} documents, \TokenizerRate{} tokens per second on one core). On the benchmark
anchor, llm.c's script scores OpenAI's released model at \AnchorLlmcAccNorm{} and ours scores
\AnchorNbAccNorm{} on the same \HellaItems{} items---identical on the length-normalised
measure, and differing by one item on the unnormalised one.

None of this proves a long run will be right. It establishes that arithmetic, data and scorer
agree before a long run starts, so anything that appears later is a property of training
rather than of those three.

\subsection{Trajectory against the reference}\label{sec:res-trajectory}

Table~\ref{tab:trajectory} lists held-out loss at every step where both runs were measured.
Agreement is close and one-sided: \numbat{} sits below the reference at all \NPaired{} points,
by \MeanDevAbs{} on average, ranging from \MinDevAbs{} at the closest point to
\MaxDevAbs{} at the widest.

\input{table_trajectory}

\input{table_trajectory_two}

That reference stopped early. The second one, run to answer a different question
(\S\ref{sec:refcross}), kept going --- and Table~\ref{tab:trajectory2} compares against it
over \NPairedTwo{} evaluations reaching step \PairedTwoMaxStep{}, three times the range.
The picture holds and sharpens: \numbat{} is below at \NBelowTwo{} of \NPairedTwo{} points,
by \MeanDevTwoAbs{} on average.

\textbf{And the gap narrows as the run proceeds}, from \WidestTwoAbs{} early to
\NarrowestTwoAbs{} by step \PairedTwoMaxStep{}. That is what the batch difference predicts.
This reference takes llm.c's 524,288-token step against \numbat{}'s \TokensPerStep{}, so at
equal step count \numbat{} has seen 0.195\% more data, an advantage largest at the
start, when the curve is steepest, and progressively less as it flattens. A framework
advantage would not behave that way; a data-position advantage does.

One-sidedness deserves comment rather than celebration. A consistent sign across eight points
is unlikely to be noise, and the most probable cause is the batch difference of
\S\ref{sec:geometry}: our step consumes \BatchDiffPct{} more tokens, so at equal step count we
have seen slightly more data. We report the gap as agreement within a narrow band, not as an
improvement.

This run has since finished all \TotalSteps{} steps and \TokensSoFar{} tokens, with held-out
loss \LastValLoss{} and HellaSwag \LastHella{}. Its endpoint against llm.c's published
\TargetValLoss{} and \TargetHella{}, and the same schedule repeated through the C~ABI, are in
\S\ref{sec:endpoint}.

\subsection{How reproducible is the reference itself?}\label{sec:refcross}

A same-machine reference has an uncomfortable property: it is one run. Every difference
between it and the framework under test is then attributed to the framework, including
whatever is peculiar to how the reference was run.

We therefore ran a second reference on a different machine---two uncapped cards instead of
three capped ones, and llm.c's exact \LlmcTokensPerStep{}-token batch, which two cards can
reach and three cannot. Across \NRefCross{} common evaluation points the two land within
\RefCrossDelta{} at step \RefCrossStep{} (\RefLabOne{} against \RefLabTwo{}, a relative
difference of \RefCrossPct{}).

This is a small number and it is the point. It bounds how much of any framework gap is method:
a difference of \MeanDevAbs{} between \numbat{} and reference is an order of magnitude larger
than a difference of \RefCrossDelta{} between two runs of that reference, so the former is not
explained by the latter. It also shows that llm.c's trajectory is not sensitive to card count
or to a fractional change in batch size, which is what makes comparison by step defensible.

\subsection{Speed}\label{sec:speed}

\input{table_sdk}

Table~\ref{tab:sdk} is the head-to-head. All three arms ran on one card in one session with
the order rotated, which matters more than it sounds: a card's clocks settle as it warms and
every stack on it slows together, so two runs minutes apart measure the card. \numbat{}'s SDK
reaches \SdkTps{} tokens per second against PyTorch's \SdkPtTps{}---\SdkRatio{}---and is ahead
in every round and in every ordering, with each arm spreading under 0.8\%. On three cards of
the same box the margin is \SdkThreeRatio{} (\SdkThreeTps{} against \SdkThreePt{}).

The extraction is deliberately not symmetric, and not in \numbat{}'s favour. The framework
trainer reports a rolling median reaching back to its sixth step, while the SDK and PyTorch
medians cover only the last fifteen of thirty; a card is cooler early, so the arm we are least
concerned to flatter is the one that gets the help.

\input{table_endtoend}

Table~\ref{tab:endtoend} places that alongside the earlier two-arm measurements, which are
each from their own session and are reported for the range of hardware they cover rather than
for fine comparison between rows. Every row is the whole loop at the configuration a real run
uses---production batch, mixed precision, data loading, evaluation and all---because a
forward-and-backward microbenchmark at a smaller batch is a different measurement that happens
to share units. PyTorch runs llm.c's own configuration with
\texttt{torch.compile}~\citep{ansel2024pytorch2}.

Results depend on how many micro-steps the recipe runs to one optimizer step. On a smaller
card---one GTX-class consumer GPU, batch \SugarBatch{}---\numbat{} reaches \SugarNb{} tokens
per second against PyTorch's \SugarPt{} and llm.c's own CUDA trainer at \SugarLlmc{}:
\SugarRatio{} and \SugarLlmcRatio{} respectively, so ahead of both.

\paragraph{Data parallelism.} One process per rank, which is the shape \texttt{torchrun} uses
and for the same reason: a single host thread submitting several devices in turn cannot
overlap them. Measured that way first, three cards gave 41,120 tokens per second against one
card's 66,558---slower in aggregate than a single card at the same per-card batch, because the
cards were taking turns. As processes they give \SdkScaleThree{} against \SdkScaleOne{}, a
speedup of \SdkSpeedup{} and an efficiency of \SdkScaleEff{}. Pairing each single-card leg
with the three-card leg beside it in time, rather than taking medians across the session,
gives \SdkPairedMedian{}; the three-card legs spread 0.6\% and the single-card legs 5\%, so
the pairing is quoted as well as the median. PyTorch scales the same three cards at
2.684$\times$ against \numbat{}'s 2.673$\times$, so neither is paying for the other's
collective.

Two defects in that path are recorded here because both look healthy. Until each rank was
given its own slice of the shard, every rank walked it from zero and the per-rank losses came
out identical---which reads as a well-synchronised run and is three replicas computing one
replica's gradient. A second defect: logged loss was rank zero's shard rather than the group's mean,
which publishes a curve nobody else can reproduce and one noisier than the real one by
$\sqrt{\mathrm{world}}$.

\input{table_levers}

\paragraph{What did not work.} Table~\ref{tab:levers} reports four
optimisations tried after the SDK reached parity. Three are losses and the fourth is inside
the noise, and we report them because two of the three \emph{pay} in \numbat{}'s own framework
trainer on the same kernels---so a lever is a property of a code path and a machine, not of a
kernel. The fused residual-and-pre-norm kernel is worth $+0.27\%$ in the framework trainer and
$-0.55\%$ here, and it is bit-identical either way: a framework gate requires its residual sum
to match the composed addition exactly.

The one that is most instructive is the last. A kernel-level profile found 147
\texttt{bmul\_f32} launches a step in the SDK and none in the framework trainer---gradient
clipping applied as a full read and write of every gradient, where the other stack measures the
norm and hands the coefficient to the optimizer so the multiply rides along with a read it was
making anyway. The arithmetic predicted $+0.06\%$. Three rounds measured $-0.34\%$, with the
coefficient arm the \emph{more} repeatable of the two. A mechanism that is real, an
arithmetic that is right, and a sign that is wrong is the ordinary case rather than the
surprising one, which is why the loop terminates on a measurement and not on a profile.

That ordering has a mechanism, and it is measured rather than inferred. Two of \numbat{}'s
costs differ from the reference's in opposite directions, and the two are paid at different
rates. Its optimizer and gradient clipping are the cheaper pair, worth about
\OptimSavingMs{}~milliseconds, and that is paid once per OPTIMIZER step. Its attention is the
dearer one, and that is paid once per MICRO-step. The smaller card's recipe runs
\SugarBatch{} tokens in a single micro-step, so it keeps the whole of the optimizer saving
against one micro-step of attention. The larger card's runs \OneGpuTokens{} tokens in
\MicroSteps{}, so the same saving is divided by \MicroSteps{} while the attention cost is
not. The advantage therefore narrows as the effective batch grows, which is arithmetic about
where each cost falls rather than a hedge about the measurement.

A second cost appears only once there is more than one card, and it is what takes the
multi-card rows below parity rather than close to it: the gradient reduction, which
\numbat{} overlaps with the backward but not yet as cheaply as the reference. It is separated
from the first below, because the two are confounded in any single row and a reader given only
the totals cannot tell which is which.

It is specifically not a host-side story, and we checked rather than assumed. Both frameworks
spend above 99\% of wall time inside kernels on an unconstrained card, so neither is waiting
on the host to keep the queue full, and kernel time per step is \KernelTotalNb{} milliseconds
for \numbat{} against \KernelTotalPt{} for PyTorch. Whatever separates them is work the card
does, not work the host fails to supply.

One caution on reading the three-card row: those cards are held at \PowerCap{}~W, so it is
not a measurement of what the hardware can do. It is a measurement of two frameworks under
the same restriction, which is what a comparison needs.

Table~\ref{tab:stage} attributes that difference. Kernel time is level---\KernelTotalPt{}
against \KernelTotalNb{} milliseconds---and wall clock differs by \WallGapPct{}. Largest
matrix multiplications run the same vendor kernels and take the same time. \numbat{} is behind
on attention, where the reference runs FlashAttention-2~\citep{dao2023flash2}, on
cross-entropy, on elementwise work that a compiler fuses into neighbours and \numbat{}
launches separately, and on optimizer updates applied per tensor rather than in one fused
call. It is ahead on embedding gradients.

\input{table_stage}

Two further findings concern measurement rather than either framework.

\textit{Reduction across cards is what turns a win into a loss.} PyTorch overlaps gradient
reduction with backward computation; \numbat{} now does too, driven by a hook that fires where
a mixed-precision node finalises a parameter gradient. It is not yet as cheap. Scaling is the
measurement that shows it: going from one card to two on the same box and recipe, \numbat{}
sustains \ScaleNb{} of twice its single-card rate and PyTorch \ScalePt{} of twice its own. That
gap of about a point is the whole of the difference between those two rows---the framework
trainer leads by \OneGpuRatio{} on one card of that box and trails at \TwoGpuRatio{} on
two---so the second card, not the per-card arithmetic, is where the two-card row is decided.
Neither box has a card-to-card link, so both collectives run through host memory and the
comparison is between two implementations of the same disadvantage.

That is not a standing property of the stack, and the three-card measurement is how we know.
Driven through the C ABI with one process per rank, \numbat{} scales one card to three at
2.673$\times$ against PyTorch's 2.684$\times$ on the same box and recipe---eleven thousandths
apart---and holds \SdkThreeRatio{} of PyTorch in aggregate. A collective that was a point
behind on two cards of one box is at parity on three of another, which says the cost is in a
particular reduction path rather than in the idea of reducing.

\textit{An instrument can hide the cost it was built to find.} Our monitoring posted progress
from inside the training thread, twice every ten steps, each post a fresh network handshake of
about a second. Both posts sat outside the region the trainer timed, so its own meter read
\MeterS{}~s per step while wall clock read \ClockS{}. Moving those posts to a background thread
took the same machine from \MonitorBefore{} to \MonitorAfter{} tokens per second,
\MonitorGain{}, and closed the gap between meter and clock to \ClockAfterS{}.

Neither of those, nor the remaining kernel gaps, explains why three capped cards sustain
\ThreeGpuNb{} tokens per second. That is the power cap: all three report a software power
limit continuously and hold between \ClockLo{} and \ClockHi{}~MHz against a \ClockMax{}~MHz
maximum, and data-parallel training runs at the slowest card's pace.

\subsection{Endpoint}\label{sec:endpoint}

The run completed its full schedule of \TotalSteps{} steps and \TokensSoFar{} tokens. Held-out
loss finished at \LastValLoss{} and HellaSwag at \LastHella{}, against llm.c's published
\TargetValLoss{} and \TargetHella{}. No optimizer step was skipped for a bad gradient at any
point in the run.
Both numbers sit slightly on the better side of the published pair, and that margin is small
enough to need a caveat rather than a claim. Our optimizer step takes \TokensPerStep{} tokens
where llm.c's takes \LlmcTokensPerStep{}, a difference of \BatchDiffPct{}, so across an equal
number of steps this run saw a little more data than the run it is compared against. That
alone is the most likely explanation for a margin of this size, and it is the same explanation
\S\ref{sec:res-trajectory} gives for the trajectory sitting a little below the reference
throughout. We read the endpoint as reproducing llm.c's result, not as improving on it.

What the endpoint does support is narrower and is the claim this paper makes: a full
\TokensSoFar{}-token pretraining run, on hardware and in a framework unrelated to the
reference, lands on the reference's published quality on both of its own measures, having run
to completion without divergence, without a skipped step, and without intervention.

\paragraph{The same run, driven through the C ABI.} We then repeated the schedule end to end
through the C~ABI rather than the framework's own Zig API: the same recipe file, the same
\SdkLastStep{} steps and the same corpus, with every tensor operation, the optimizer, the
data loader and the collective reached through \texttt{nb\_*} calls. It finished at held-out
loss \SdkLastValLoss{} and HellaSwag \SdkLastHella{} (\SdkLastHellaAcc{} unnormalised, over
\SdkHellaItems{} items), which is \SdkVsTreeLoss{} loss and \SdkVsTreeHella{} HellaSwag
against the framework's own endpoint above, and \SdkVsLlmcLoss{} and \SdkVsLlmcHella{}
against llm.c's published figures.

Two caveats belong with those numbers. HellaSwag was not scored during the run -- the binding's
trainer has no scorer -- so the endpoint figure was measured afterwards from the last checkpoint
with the framework's evaluator, the same one anchored at \AnchorNbAccNorm{} on the released
OpenAI checkpoint. That checkpoint is step~18{,}750 rather than 18{,}865, because the
checkpoint cadence is 250 and the trainer wrote no final weights; held-out loss is
\SdkLastValLoss{} at both steps and the learning rate has decayed to zero by then, so the
distinction is bookkeeping rather than a different model, but it is the checkpoint that was
scored and we say so.

Both of those are properties of this run rather than of the stack, and both have since been
closed in the trainer: it writes a checkpoint at the final step whatever the cadence, and a
resumed run now continues its metric log instead of truncating it---the framework's logger
had always supported continuing a log, and the C~ABI had no door onto it, so every
binding's resume destroyed its own history. We report the run as it happened, with the
curve as we recovered it; a reader should not infer that either limitation is still
there.

What the pair supports is narrow: a difference of
\SdkVsTreeLoss{} in loss over 9.91~B tokens is smaller than the run-to-run variation we
measure in \S\ref{sec:refcross}, so the binding is not a reduced surface that trades accuracy
for portability. It is the same computation reached a different way.

\subsection{Clinical adaptation}\label{sec:clinical}

\input{table_clinical}

Over one epoch of \ClinSteps{} steps for \numbat{} and \ClinStepsPt{} for the
reference, best held-out loss is \ClinNbLoss{} against \ClinPtLoss{}, a difference of
\ClinLossGap{} (\ClinLossGapPct{}) in \numbat{}'s favour. Perplexity is \ClinNbPpl{}
against \ClinPtPpl{}. Both legs score the same capped \ClinScored{}-example prefix of
the same held-out split and both reach their best at step 3{,}500, so the two figures
are one measurement at one step rather than two similar ones; the gate refuses the
comparison when the scored counts differ.

The step counts differ by one because one epoch of 168{,}473 examples at
\ClinPerStep{} an optimizer step is 3{,}509.85 steps, and the two legs round it
differently---\numbat{} drops the partial step, the reference harness keeps it. It is
0.03\% of the schedule and we report it rather than rounding it away. It also explains
the evaluation counts: \ClinEvals{} against \ClinEvalsPt{}, because the reference also
evaluates on its final step and \numbat{}'s last step is not a multiple of the
evaluation cadence.

\textbf{On speed, this pair is not the measurement.} The two legs' wall clocks were
3.00 and 2.43 hours, and dividing them would be wrong: the reference leg shared the
machine with an unrelated compile for part of its run---its step time moved from
2.74\,s to 2.36\,s when that finished, about 14\%---while the \numbat{} leg had the box
to itself. What the machine is doing is part of a wall clock, so a pair measured under
different conditions cannot be divided. The speed comparison is the interleaved
benchmark of \S\ref{sec:speed}: both arms at this recipe's own production geometry,
each alone on the box, warmup discarded, \numbat{} at \ClinNbMs{}\,ms an optimizer step
against \ClinPtMs{}, or \ClinSpeedMean{}. We quote the ratio of MEANS because an
epoch's wall clock is a mean; by medians it is \ClinSpeedMedian{}, which describes the
steady-state step rather than the run.

\subsection{Operational record}\label{sec:operational}

Across the clinical epoch the gate issued \ClinGateVerdicts{} verdicts: \ClinGateOK{} normal
and \ClinGateNoData{} before a first evaluation existed. No alert, no termination. That is the
intended outcome, and distinct from an absent control: a sidecar with
authority to stop the run, which never needed to use it, is a different situation from no
sidecar.

\section{Availability}\label{sec:availability}

Two public repositories accompany this paper, both under the Apache License~2.0, and
both carry training logs as well as weights so that a reader can audit a claim rather
than accept it.

\paragraph{Pretrained weights.} The GPT-2 endpoint of \S\ref{sec:endpoint} is released
as a single \texttt{safetensors} file of \NParams{} parameters, with a
\texttt{config.json} giving the architecture and an \texttt{eval\_curve.json} holding all
76 held-out evaluations.\footnote{\url{https://huggingface.co/cloudkites/gpt2-124m-fineweb-edu}}
Tensor names follow llm.c's flat convention rather than HuggingFace's and the model card
says how to map between them; the vocabulary is padded to 50{,}304 while only the first
50{,}257 rows are scored, which a reader recomputing perplexity needs to know.

\paragraph{Clinical adapter.} The LoRA adapter of \S\ref{sec:clinical} is released with
its \ClinTrainable{} trainable parameters, a PEFT-shaped configuration that loads against
the released Qwen3-0.6B, and its own evaluation
curve.\footnote{\url{https://huggingface.co/cloudkites/qwen3-0.6b-clinical-lora}} Neither
it nor the base model is a medical device, and neither is validated for clinical use.

\paragraph{Training logs, for audit and reproduction.} Each repository carries a
\texttt{logs/} directory. For the clinical adapter this is the complete per-step record
of all \ClinSteps{} optimizer steps---loss, learning rate, gradient norm, supervised
tokens and throughput---together with every verdict the supervising gate issued and
\emph{the reference leg's own curve}, so the agreement reported in
Table~\ref{tab:clinical} can be recomputed from published data rather than read from a
table. For the pretrained model the logs are the surviving per-step segment and the
run's summary; the metric writer truncated its own files on resume, so the full
evaluation curve was recovered from console output and is published separately. We state
that limitation in the repository rather than presenting a partial log as a complete one.

\paragraph{What is not released.} \numbat{} itself is not published. It is under active
development, and public access is intended once its interfaces are stable and mature
enough that a consumer can depend on them; releasing a moving surface would hand readers
something that breaks under them. The asymmetry is a matter of timing rather than
principle, and it costs a reader less than it may appear: the artefact needed to check a
claim about a trained model is the trained model and its record, both of which are here,
and none of the framework is required to load or run either set of weights. The
implementation we reproduce (llm.c) and the corpus (FineWeb-Edu, ODC-By~1.0) are already
public, so the pipeline is reconstructible from published components independently of our
implementation of it.

\section{Discussion}\label{sec:discussion}

\paragraph{What close agreement is, and is not, evidence for.} Two stacks passing through the
same sequence of held-out losses is strong evidence that they implement the same computation,
and weak evidence about whether that computation is the right one. Both could be wrong
together. \citet{knight1986nversion} established experimentally that independently written
versions do not fail independently, and nothing here escapes that. What a second
implementation buys is coverage of faults that are not common-mode, which empirically is most
of them.

\paragraph{A reference needs a control of its own.} We think \S\ref{sec:refcross} is the most
transferable part of this work. A same-machine reference is widely recommended and rarely
audited, and without a second run of that reference, framework difference and reference
peculiarity are indistinguishable. Ours cost a few hours on an otherwise idle machine and
turned a one-sided \MeanDevAbs{} gap from an unexplained observation into one with a bounded
alternative explanation.

\paragraph{Observation is not control.} A truncated reference is the normal case on a small
budget, and a band drawn past its end is vacuous for exactly the stretch where a long run
most needs watching. Separating a modelled band that can only warn from a self-referential
stall check that can stop seems to us the right shape, because it keeps termination authority
attached to a judgement that cannot be wrong about a curve it never measured.

\paragraph{Where the remaining speed gap is.} Nothing in Table~\ref{tab:stage} is mysterious,
and none of it is arithmetic. Largest matrix multiplications are at parity on the same vendor
kernels. Fusion of small operations, the optimizer step and overlap of gradient reduction with
backward work were each a measurable part of this gap and are each now closed, which leaves
attention as the whole of what is left. There the reference runs
FlashAttention-2~\citep{dao2023flash2} and \numbat{} calls a vendor fused kernel that is about
a fifth dearer at identical shapes and layout, on the forward and on the backward alike. We
wrote a replacement to test whether that is reachable; it is numerically correct and still
slower than the vendor kernel it would replace, so it stays behind a flag and the vendor path
stays the default. We report the gap unclosed rather than quoting a favourable configuration,
and note that it is now a kernel-writing problem with a measured mechanism rather than a gap
of unknown origin.

\paragraph{Why a medical stage.} Domains differ in what a quiet defect costs. A stack that
trains a general model to reference parity and is then used unexamined on clinical text
inherits none of that assurance, because the pipeline around the framework changes---how text
is rendered into a prompt, what is masked, what is scored. An earlier study found exactly
there the fault with largest effect on a trained model, outside numerical kernels
altogether~\citep{tran2026crossstack}. Running both stages under the same controls is what
lets the assurance carry.

\section{Limitations}\label{sec:limits}

\textit{Reference coverage.} Our same-machine reference spans step \RefLastMeasured{} of
\TotalSteps{}. Beyond that, comparison is against a published endpoint and an interpolated
band that can only warn. \S\ref{sec:refcross} bounds reproducibility of the reference but does
not extend its range.

\textit{One machine, one scale, one architecture.} Results are for one model size on one
generation of hardware with three cards and no direct card-to-card link. Reduction cost in
particular is a property of that interconnect, and would differ on a machine with one.

\textit{Batch geometry.} Our step is \BatchDiffPct{} larger than llm.c's, in a direction that
favours us at equal step count. We believe this explains the sign in
\S\ref{sec:res-trajectory} and we have not isolated it.

\textit{Clinical claims.} We measure held-out cross-entropy on clinical question-answer text.
That is a statement about a language model's fit to a corpus and not about clinical accuracy,
safety or usefulness. Nothing here was evaluated by a clinician, against patient outcomes, or
for the failure modes that matter in care. The adapted model is a research artefact, is not a
medical device, is not validated for clinical use, and must not inform patient care.

\textit{Speed comparisons state their conditions.} Pretraining comparisons use
\texttt{torch.compile}; the clinical reference leg does not, because its recipe does not.
Reading either number outside its configuration would be a mistake.

\section{Conclusion}\label{sec:conclusion}

We trained a GPT-2 of \NParamsM{} parameters from random initialisation on a machine-learning
stack that shares no code with the framework nearly all such work uses, and adapted a separate
model to clinical question answering with the same stack. Both stages were checked against a
reference implementation on the same machine, and both were supervised by a process able to
stop them.

Held-out loss tracked that reference closely and one-sidedly across every paired evaluation.
A second reference run on different hardware agreed with the first to \RefCrossDelta{}, which
is what lets us say the framework gap is not an artefact of how the reference was produced.
Speed was not sacrificed for it: \SdkRatio{} of a compiled PyTorch on one card and
\SdkThreeRatio{} on three, measured with every arm in one session, and \SdkSpeedup{} scaling
across those three cards. Where a margin remained we attributed it to a named mechanism and
reported it unclosed---including four optimisations that measured as losses, and one published
figure we had to withdraw because it compared two sessions rather than two stacks.

We take the combined result as evidence that an independent implementation can carry a
language model through its whole life---from noise, through pretraining, into a specialist
domain---and be checked at each stage rather than trusted. Whether that is worth a second
implementation's cost is a judgement each project makes. What this study supplies is a measured
estimate of the cost, and a description of the controls that make such a run answerable.

\section*{Availability}

Every number in this paper is produced by a generator from files under \texttt{data/}: the
run's own evaluation curve, both reference trajectories, apparatus description, and the
measurements quoted from gates. No framework source is included in those artefacts.

\section*{Acknowledgements}

We thank the authors of llm.c for a reproduction written to be checked against, and the
FineWeb authors for a corpus documented well enough to prepare independently.

\bibliographystyle{plainnat}
\bibliography{references}

\end{document}

%% file: results.tex
\newcommand{\NLayer}{12\xspace}
\newcommand{\NHead}{12\xspace}
\newcommand{\NEmbd}{768\xspace}
\newcommand{\BlockSize}{1{,}024\xspace}
\newcommand{\VocabPadded}{50{,}304\xspace}
\newcommand{\VocabReal}{50{,}257\xspace}
\newcommand{\NParams}{124{,}439{,}808\xspace}
\newcommand{\NParamsM}{124.4\,M\xspace}
\newcommand{\PeakLR}{$6\times10^{-4}$\xspace}
\newcommand{\WarmupSteps}{700\xspace}
\newcommand{\WeightDecay}{0.1\xspace}
\newcommand{\GradClip}{1.0\xspace}
\newcommand{\AdamBetas}{$(0.9, 0.95)$\xspace}
\newcommand{\Seed}{1{,}337\xspace}
\newcommand{\WorldSize}{3\xspace}
\newcommand{\PerDeviceBatch}{19\xspace}
\newcommand{\GradAccum}{9\xspace}
\newcommand{\SeqLen}{1{,}024\xspace}
\newcommand{\TokensPerStep}{525{,}312\xspace}
\newcommand{\LlmcTokensPerStep}{524{,}288\xspace}
\newcommand{\BatchDiffPct}{0.195\%\xspace}
\newcommand{\TotalSteps}{18{,}865\xspace}
\newcommand{\TokensTotal}{9.91\,B\xspace}
\newcommand{\LlmcTokensTotal}{9.89\,B\xspace}
\newcommand{\ValEvery}{250\xspace}
\newcommand{\ValTokens}{10{,}506{,}240\xspace}
\newcommand{\LlmcValTokens}{10{,}485{,}760\xspace}
\newcommand{\TargetValLoss}{3.29\xspace}
\newcommand{\TargetHella}{0.299\xspace}

\newcommand{\CheckLogit}{$1.465\times10^{-3}$\xspace}
\newcommand{\CheckLoss}{$7.153\times10^{-6}$\xspace}
\newcommand{\CheckGrad}{$2.408\times10^{-5}$\xspace}
\newcommand{\CheckSteps}{$1.31\times10^{-4}$\xspace}
\newcommand{\CheckTol}{$10^{-2}$\xspace}
\newcommand{\CheckWorstTensor}{\texttt{h.10.ln\_2.weight}\xspace}
\newcommand{\CheckParams}{124{,}439{,}808\xspace}
\newcommand{\NGradTensors}{148\xspace}
\newcommand{\ShardTokens}{100{,}000{,}000\xspace}
\newcommand{\ShardBytes}{200{,}000{,}000\xspace}
\newcommand{\ShardDocs}{143{,}468\xspace}
\newcommand{\TokenizerRate}{5.7\,M\xspace}
\newcommand{\HellaItems}{10{,}042\xspace}
\newcommand{\AnchorLlmcAccNorm}{0.2955\xspace}

\newcommand{\AnchorNbAccNorm}{0.2955\xspace}

\newcommand{\LastValLoss}{3.2588\xspace}
\newcommand{\LastHella}{0.3053\xspace}

\newcommand{\TokensSoFar}{9.91\,B\xspace}

\newcommand{\SdkLastStep}{18{,}865\xspace}
\newcommand{\SdkLastValLoss}{3.2598\xspace}
\newcommand{\SdkLastHella}{0.3036\xspace}
\newcommand{\SdkLastHellaAcc}{0.2895\xspace}
\newcommand{\SdkHellaItems}{10{,}042\xspace}
\newcommand{\SdkVsTreeLoss}{+0.0010\xspace}
\newcommand{\SdkVsTreeHella}{-0.0017\xspace}
\newcommand{\SdkVsLlmcLoss}{-0.0302\xspace}
\newcommand{\SdkVsLlmcHella}{+0.0046\xspace}

\newcommand{\NPaired}{8\xspace}
\newcommand{\RefLastMeasured}{2{,}000\xspace}

\newcommand{\MeanDevAbs}{0.0608\xspace}
\newcommand{\MinDevAbs}{0.0412\xspace}
\newcommand{\MaxDevAbs}{0.1084\xspace}
\newcommand{\RefCrossStep}{2{,}000\xspace}
\newcommand{\RefLabOne}{3.8525\xspace}
\newcommand{\RefLabTwo}{3.8560\xspace}
\newcommand{\RefCrossDelta}{0.0035\xspace}
\newcommand{\RefCrossPct}{0.09\%\xspace}
\newcommand{\NRefCross}{9\xspace}
\newcommand{\NPairedTwo}{25\xspace}
\newcommand{\PairedTwoMaxStep}{6{,}250\xspace}
\newcommand{\MeanDevTwoAbs}{0.0439\xspace}
\newcommand{\NBelowTwo}{25\xspace}
\newcommand{\WidestTwoAbs}{0.1231\xspace}
\newcommand{\NarrowestTwoAbs}{0.0274\xspace}
\newcommand{\SugarBatch}{4{,}096\xspace}
\newcommand{\SugarNb}{23{,}045\xspace}
\newcommand{\SugarPt}{21{,}526\xspace}
\newcommand{\SugarLlmc}{20{,}507\xspace}
\newcommand{\SugarRatio}{1.071$\times$\xspace}
\newcommand{\SugarLlmcRatio}{1.124$\times$\xspace}

\newcommand{\OneGpuRatio}{1.008$\times$\xspace}

\newcommand{\TwoGpuRatio}{0.995$\times$\xspace}
\newcommand{\ThreeGpuNb}{115{,}938\xspace}
\newcommand{\SdkTps}{43{,}374\xspace}

\newcommand{\SdkPtTps}{41{,}202\xspace}
\newcommand{\SdkRatio}{1.053$\times$\xspace}

\newcommand{\SdkThreeTps}{115{,}938\xspace}
\newcommand{\SdkThreePt}{110{,}589\xspace}
\newcommand{\SdkThreeRatio}{1.048$\times$\xspace}
\newcommand{\SdkScaleOne}{43{,}052\xspace}
\newcommand{\SdkScaleThree}{119{,}215\xspace}
\newcommand{\SdkSpeedup}{2.769$\times$\xspace}
\newcommand{\SdkScaleEff}{92.3\%\xspace}
\newcommand{\SdkPairedMedian}{2.754$\times$\xspace}
\newcommand{\OptimSavingMs}{5\xspace}
\newcommand{\OneGpuTokens}{262{,}144\xspace}
\newcommand{\MicroSteps}{sixteen\xspace}
\newcommand{\ScaleNb}{92.5\%\xspace}
\newcommand{\ScalePt}{93.7\%\xspace}
\newcommand{\KernelTotalPt}{291\xspace}
\newcommand{\KernelTotalNb}{292\xspace}

\newcommand{\WallGapPct}{3\%\xspace}
\newcommand{\MonitorBefore}{101{,}560\xspace}
\newcommand{\MonitorAfter}{108{,}814\xspace}
\newcommand{\MonitorGain}{7.1\%\xspace}
\newcommand{\MeterS}{4.80\xspace}
\newcommand{\ClockS}{5.17\xspace}
\newcommand{\ClockAfterS}{4.83\xspace}
\newcommand{\ClockLo}{1{,}155\xspace}
\newcommand{\ClockHi}{1{,}335\xspace}
\newcommand{\ClockMax}{2{,}100\xspace}
\newcommand{\PowerCap}{200\xspace}

\newcommand{\ClinSteps}{3{,}509\xspace}
\newcommand{\ClinStepsPt}{3{,}510\xspace}
\newcommand{\ClinEvals}{14\xspace}
\newcommand{\ClinEvalsPt}{15\xspace}
\newcommand{\ClinPerStep}{48\xspace}

\newcommand{\ClinSpeedMean}{1.0263$\times$\xspace}
\newcommand{\ClinSpeedMedian}{1.0399$\times$\xspace}
\newcommand{\ClinNbMs}{2{,}243\xspace}
\newcommand{\ClinPtMs}{2{,}302\xspace}
\newcommand{\ClinTrain}{168{,}473\xspace}
\newcommand{\ClinEval}{3{,}440\xspace}
\newcommand{\ClinScored}{512\xspace}
\newcommand{\ClinRecordsIn}{194{,}000\xspace}
\newcommand{\ClinAccounted}{100.0\%\xspace}
\newcommand{\ClinTrainable}{10{,}092{,}544\xspace}
\newcommand{\ClinRank}{16\xspace}
\newcommand{\ClinAlpha}{32\xspace}
\newcommand{\ClinNbLoss}{2.1899\xspace}
\newcommand{\ClinPtLoss}{2.1941\xspace}
\newcommand{\ClinNbPpl}{8.935\xspace}
\newcommand{\ClinPtPpl}{8.972\xspace}

\newcommand{\ClinLossGap}{0.0042\xspace}
\newcommand{\ClinLossGapPct}{0.19\%\xspace}
\newcommand{\ClinGateVerdicts}{146\xspace}
\newcommand{\ClinGateOK}{134\xspace}
\newcommand{\ClinGateNoData}{12\xspace}
\newcommand{\ClinGatePoll}{1\xspace}
\newcommand{\GateBandPct}{5\%\xspace}

%% file: fig_architecture.tex
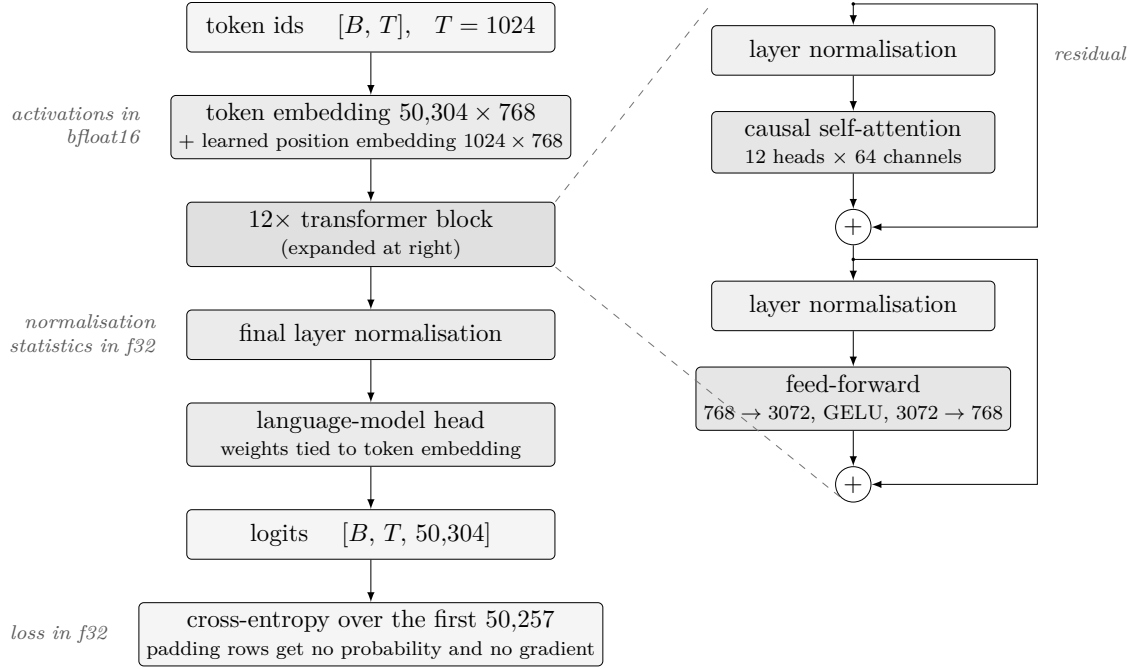
\begin{figure}[t]
\centering
\resizebox{\textwidth}{!}{%
\begin{tikzpicture}[
  font=\small,
  node distance=0pt,
  box/.style   = {draw, rounded corners=2pt, minimum width=46mm, minimum height=7mm,
                  align=center, inner sep=3pt},
  wide/.style  = {box, minimum width=52mm},
  op/.style    = {draw, circle, inner sep=0pt, minimum size=5mm},
  lbl/.style   = {font=\scriptsize\itshape, text=black!65},
  ar/.style    = {-latex, thin},
  blk/.style   = {draw, dashed, rounded corners=3pt, inner sep=5pt},
]

\node[wide, fill=black!4] (tok) {token ids \quad $[B,\,T]$, \; $T=1024$};
\node[wide, below=6mm of tok, fill=black!8] (emb)
  {token embedding $50{,}304\times768$\\[-1pt]
   {\scriptsize$+$ learned position embedding $1024\times768$}};
\node[wide, below=6mm of emb, fill=black!12] (blocks)
  {$12\times$ transformer block\\[-1pt]{\scriptsize(expanded at right)}};
\node[wide, below=6mm of blocks, fill=black!8] (lnf) {final layer normalisation};
\node[wide, below=6mm of lnf, fill=black!8] (head)
  {language-model head\\[-1pt]{\scriptsize weights tied to token embedding}};
\node[wide, below=6mm of head, fill=black!4] (logits) {logits \quad $[B,\,T,\,50{,}304]$};
\node[wide, below=6mm of logits, fill=black!4] (loss)
  {cross-entropy over the first $50{,}257$\\[-1pt]
   {\scriptsize padding rows get no probability and no gradient}};

\foreach \a/\b in {tok/emb, emb/blocks, blocks/lnf, lnf/head, head/logits, logits/loss}
  \draw[ar] (\a) -- (\b);

\node[box, right=22mm of blocks.east, yshift=26mm, minimum width=40mm, fill=black!6] (ln1)
  {layer normalisation};
\node[box, below=5mm of ln1, minimum width=40mm, fill=black!10] (att)
  {causal self-attention\\[-1pt]{\scriptsize12 heads $\times$ 64 channels}};
\node[op, below=5mm of att] (add1) {$+$};
\node[box, below=5mm of add1, minimum width=40mm, fill=black!6] (ln2)
  {layer normalisation};
\node[box, below=5mm of ln2, minimum width=40mm, fill=black!10] (mlp)
  {feed-forward\\[-1pt]{\scriptsize $768\rightarrow3072$, GELU, $3072\rightarrow768$}};
\node[op, below=5mm of mlp] (add2) {$+$};

\draw[ar] (ln1) -- (att);
\draw[ar] (att) -- (add1);
\draw[ar] (add1) -- (ln2);
\draw[ar] (ln2) -- (mlp);
\draw[ar] (mlp) -- (add2);

%
\coordinate (r1) at ($(ln1.north)+(0,3mm)$);
\draw[ar] (r1) -- (ln1.north);
\fill (r1) circle (0.7pt);
\draw[ar] (r1) -- ++(26mm,0) |- (add1.east);
\coordinate (r2) at ($(add1.south)+(0,-2mm)$);
\fill (r2) circle (0.7pt);
\draw[ar] (r2) -- ++(26mm,0) |- (add2.east);
\node[lbl, anchor=west] at ($(r1)+(27mm,-7mm)$) {residual};

\draw[dashed, thin, black!55] (blocks.north east) -- ($(ln1.north west)+(0,3mm)$);
\draw[dashed, thin, black!55] (blocks.south east) -- (add2.south west);

\node[lbl, anchor=east, text width=30mm, align=right] at ($(emb.west)+(-3mm,0)$)
  {activations in\\bfloat16};
\node[lbl, anchor=east, text width=30mm, align=right] at ($(lnf.west)+(-3mm,0)$)
  {normalisation\\statistics in f32};
\node[lbl, anchor=east, text width=30mm, align=right] at ($(loss.west)+(-3mm,0)$)
  {loss in f32};

\end{tikzpicture}%
}
\caption{GPT-2 (124M) as trained here: \NLayer{} blocks, \NHead{} attention heads,
\NEmbd{} channels, context \BlockSize{}. Vocabulary is padded to \VocabPadded{} so that
matrix shapes are multiples of 128, while the softmax covers only the \VocabReal{} real
tokens, leaving padding rows without probability or gradient. Master weights, gradients and
optimizer state stay in 32-bit throughout; only activations are half precision.}
\label{fig:architecture}
\end{figure}

%% file: fig_pipeline.tex
\begin{figure}[t]
\centering
\begin{tikzpicture}[
  font=\small,
  src/.style   = {draw, rounded corners=2pt, minimum height=8mm, align=center, inner sep=4pt, fill=black!6},
  proc/.style  = {draw, minimum height=8mm, align=center, inner sep=4pt, fill=black!10},
  shard/.style = {draw, minimum width=9mm, minimum height=6mm, inner sep=1pt, font=\scriptsize},
  use/.style   = {draw, dashed, rounded corners=2pt, align=left, inner sep=4pt, font=\scriptsize},
  ar/.style    = {-latex, thin},
  note/.style  = {font=\scriptsize\itshape, text=black!65, align=left},
]

\node[src] (fw) {FineWeb, ten-billion-token sample\\[-1pt]{\scriptsize web documents}};
\node[proc, below=6mm of fw] (tokenize)
  {GPT-2 byte-level encoding\\[-1pt]
   {\scriptsize each document preceded by an end-of-text marker}};
\node[proc, below=6mm of tokenize] (concat)
  {concatenated in dataset order, cut into shards of exactly $10^{8}$ tokens\\[-1pt]
   {\scriptsize a document crossing a boundary continues in the next shard}};
\draw[ar] (fw) -- (tokenize);
\draw[ar] (tokenize) -- (concat);

\node[shard, fill=black!25, below=8mm of concat.south, xshift=-46mm] (s0) {0};
\node[shard, right=1.2mm of s0] (s1) {1};
\node[shard, right=1.2mm of s1] (s2) {2};
\node[shard, right=1.2mm of s2] (s3) {3};
\node[right=1.2mm of s3, font=\scriptsize] (dots) {$\cdots$};
\node[shard, right=1.2mm of dots] (s103) {103};
\draw[ar] (concat.south) -- ++(0,-3mm) -| (s2.north);

\node[note, anchor=east] at ($(s0.west)+(-2mm,0)$) {held out};
\node[note, anchor=west] at ($(s103.east)+(2mm,0)$) {training,\\read in order};

\node[use, below=13mm of s0.south, anchor=north west, xshift=-2mm, text width=40mm] (val)
  {\textbf{Validation}\\
   first \ValTokens{} tokens of shard 0\\
   every \ValEvery{} steps\\
   averaged across the three ranks\\
   \emph{never trained on}};

\node[use, right=5mm of val.north east, anchor=north west, text width=46mm] (train)
  {\textbf{Training}\\
   $3$ ranks $\times$ \PerDeviceBatch{} sequences $\times$ \SeqLen{} tokens\\
   $\times$ \GradAccum{} accumulation steps\\
   $=$ \TokensPerStep{} tokens per step\\
   $\times$ \TotalSteps{} steps $=$ \TokensTotal{} tokens};

\node[use, right=5mm of train.north east, anchor=north west, text width=40mm] (hs)
  {\textbf{HellaSwag}\\
   \HellaItems{} items, a separate corpus\\
   every \ValEvery{} steps\\
   four endings scored per item,\\
   length-normalised\\
   \emph{never trained on}};

\draw[ar] (s0.south) -- ++(0,-4mm) -| ($(val.north)+(0,0)$);
\draw[ar] (s2.south) -- ++(0,-8mm) -| ($(train.north)+(0,0)$);

\node[src, above=6mm of hs.north, anchor=south, minimum height=6mm, font=\scriptsize]
  (hsrc) {HellaSwag validation set};
\draw[ar] (hsrc) -- (hs.north);

\node[note, below=4mm of train.south, anchor=north, text width=105mm, align=center]
  {Held-out data is a contiguous prefix of the corpus, so the split is by construction
   rather than by sampling, and no training token can appear in it.};

\end{tikzpicture}
\caption{From corpus to the three things it is used for. Preparation reproduces llm.c's
shard layout exactly, which is what allows this run to read llm.c's own shards; numbat's
independent implementation of the same preparation was checked against it byte for byte
(\S\ref{sec:data}). Validation and HellaSwag are scored at the same cadence, and neither
is ever trained on.}
\label{fig:pipeline}
\end{figure}
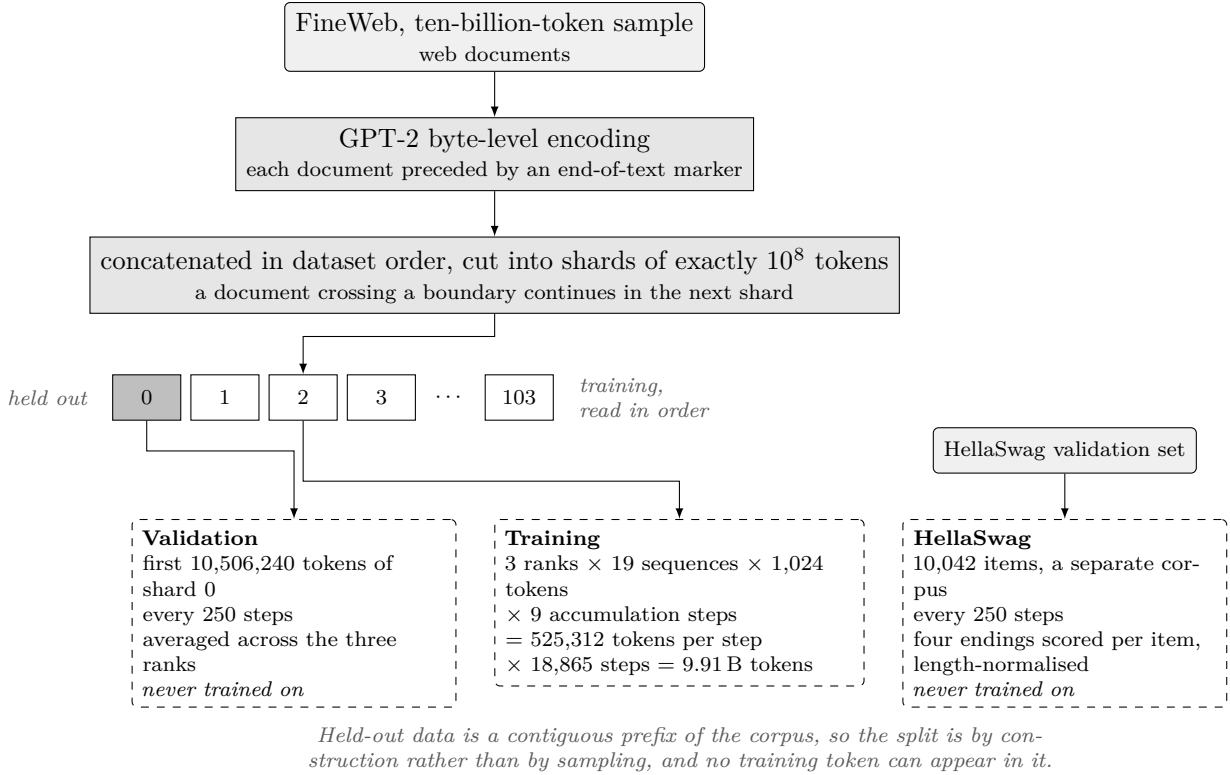

%% file: table_trajectory.tex
\begin{table}[t]
\caption{Held-out cross-entropy at every step where both runs were measured.
The reference is llm.c's own PyTorch trainer on the same three cards, stopped at
step 2{,}000 when those cards were needed for the run it gates. Lower is better;
a negative difference is numbat below the reference.}
\label{tab:trajectory}
\centering\small
\begin{tabular}{@{}rrrr@{}}
\toprule
Step & \numbat{} & llm.c reference & Difference \\
\midrule
250 & 6.1086 & 6.1509 & -0.0423 \\
500 & 5.3120 & 5.3979 & -0.0859 \\
750 & 4.6136 & 4.7220 & -0.1084 \\
1{,}000 & 4.2607 & 4.3116 & -0.0509 \\
1{,}250 & 4.0655 & 4.1237 & -0.0582 \\
1{,}500 & 3.9554 & 3.9993 & -0.0439 \\
1{,}750 & 3.8703 & 3.9258 & -0.0555 \\
2{,}000 & 3.8113 & 3.8525 & -0.0412 \\
\bottomrule
\end{tabular}
\end{table}

%% file: table_trajectory_two.tex
\begin{table}[t]
\caption{The same comparison against the SECOND reference, which ran long enough to
cover 6{,}250 steps rather than 2{,}000. Its step is llm.c's
524{,}288 tokens against numbat's 525{,}312, so numbat has seen 0.195\% more data at equal
step count -- which is the direction of the gap, and consistent with it narrowing as the
run proceeds and an early data advantage matters less.}
\label{tab:trajectory2}
\centering\small
\begin{tabular}{@{}rrrr@{}}
\toprule
Step & \numbat{} & llm.c reference & Difference \\
\midrule
250 & 6.1086 & 6.1562 & -0.0476 \\
500 & 5.3120 & 5.4032 & -0.0912 \\
750 & 4.6136 & 4.7367 & -0.1231 \\
1{,}000 & 4.2607 & 4.3245 & -0.0638 \\
1{,}250 & 4.0655 & 4.1297 & -0.0642 \\
1{,}500 & 3.9554 & 4.0010 & -0.0456 \\
1{,}750 & 3.8703 & 3.9206 & -0.0503 \\
2{,}000 & 3.8113 & 3.8560 & -0.0447 \\
2{,}250 & 3.7663 & 3.8038 & -0.0375 \\
2{,}500 & 3.7252 & 3.7630 & -0.0378 \\
2{,}750 & 3.6937 & 3.7321 & -0.0384 \\
3{,}000 & 3.6644 & 3.7029 & -0.0385 \\
3{,}250 & 3.6383 & 3.6727 & -0.0344 \\
3{,}500 & 3.6176 & 3.6516 & -0.0340 \\
3{,}750 & 3.5968 & 3.6359 & -0.0391 \\
4{,}000 & 3.5808 & 3.6136 & -0.0328 \\
4{,}250 & 3.5647 & 3.5993 & -0.0346 \\
4{,}500 & 3.5496 & 3.5859 & -0.0363 \\
4{,}750 & 3.5370 & 3.5657 & -0.0287 \\
5{,}000 & 3.5242 & 3.5535 & -0.0293 \\
5{,}250 & 3.5128 & 3.5402 & -0.0274 \\
5{,}500 & 3.5015 & 3.5346 & -0.0331 \\
5{,}750 & 3.4912 & 3.5203 & -0.0291 \\
6{,}000 & 3.4815 & 3.5105 & -0.0290 \\
6{,}250 & 3.4719 & 3.5000 & -0.0281 \\
\bottomrule
\end{tabular}
\end{table}

%% file: table_sdk.tex
\begin{table}[t]
\caption{The two \numbat{} front doors and PyTorch, measured on ONE card in ONE
session with the three arms rotated so none is always first onto a cold card.
B=16 x 1024 x 16 accum = 262,144 tokens/step, one RTX 3090 at a 200 W cap. This table exists because the figures it replaces did not come
from one session: an SDK row of 1.0003$\times$ and a framework row of
1.0085$\times$, taken on different machines on different days, were read as
0.8\% of overhead from crossing the C ABI. Measured side by side the SDK is
1.029$\times$ the framework trainer, not 0.992$\times$ of it.}
\label{tab:sdk}
\centering\small
\begin{tabular}{@{}lrrl@{}}
\toprule
Stack & Median tok/s & vs PyTorch & Rounds \\
\midrule
\numbat{} SDK (C ABI) & \textbf{43{,}374} & \textbf{1.053$\times$} & 43{,}695, 43{,}353, 43{,}413, 43{,}374 \\
\numbat{} framework trainer & 42{,}159 & 1.023$\times$ & 42{,}159, 42{,}124, 42{,}241, 42{,}235 \\
PyTorch 2.14 + \texttt{compile} & 41{,}202 & --- & 41{,}230, 41{,}284, 41{,}202, 41{,}171 \\
\bottomrule
\end{tabular}
\end{table}

%% file: table_endtoend.tex
\begin{table}[t]
\caption{Tokens per second for the whole training loop at the configuration a real
run uses, not a forward-and-backward microbenchmark. PyTorch is llm.c's own
configuration: bf16, FlashAttention and \texttt{torch.compile}. Every row is the two
sides run ALTERNATELY on one box, because a card's clocks settle as it warms and
both frameworks slow together, so two runs minutes apart measure the card.
\emph{Stack} is which of \numbat{}'s two trainers produced the row:
\emph{framework} is written against \numbat{}'s own API, \emph{SDK} is composed
entirely from the C~ABI that ships to consumers (\S\ref{sec:stack}). They are the
same core behind two doors, so a row from one is not evidence about the other.}
\label{tab:endtoend}
\centering\small
\begin{tabularx}{\textwidth}{@{}llXrrr@{}}
\toprule
Box, cards & Stack & Configuration & PyTorch & \numbat{} & Ratio \\
\midrule
Box 1, 1 & SDK & B=16 x 1024 x 16 accum = 262,144 tokens/step & 41{,}202 & 43{,}374 & 1.053$\times$ \\
Box 1, 3 & SDK & B=16 x 1024 x 16 accum x 3 = 786,432 tokens/step & 110{,}589 & 115{,}938 & 1.048$\times$ \\
Box 3, 1 & framework & B=4 x 1024 x 1 accum = 4,096 tokens/step & 21{,}526 & 23{,}045 & 1.071$\times$ \\
Box 2, 1 & framework & B=16 x 1024 x 16 accum = 262,144 tokens/step & 66{,}251 & 66{,}812 & 1.008$\times$ \\
Box 2, 2 & framework & B=16 x 1024 x 16 accum x 2 = 524,288 tokens/step (llm.c's batch exactly) & 124{,}207 & 123{,}639 & 0.995$\times$ \\
Box 1, 3 & framework & B=19 x 1024 x 9 accum x 3 = 525,312 tokens/step & 113{,}502 & 112{,}869 & 0.994$\times$ \\
\bottomrule
\end{tabularx}

\vspace{2pt}
{\footnotesize\raggedright Boxes, by hardware: \textbf{Box 1} 3$\times$ NVIDIA RTX 3090, 200~W power cap; \textbf{Box 2} 2$\times$ NVIDIA RTX 3090; \textbf{Box 3} 1$\times$ NVIDIA RTX 3060.\par}
\end{table}

%% file: table_levers.tex
\begin{table}[t]
\caption{Four optimisations tried on the SDK trainer after it reached parity,
and what each measured. Three are losses. Two of them PAY in \numbat{}'s own
framework trainer on the same kernels, which is the point of reporting them: a
lever is a property of a code path and a machine, not of a kernel. Three rounds
each, one RTX 3090 at a 200\,W cap, the configurations rotated.}
\label{tab:levers}
\centering\small
\begin{tabularx}{\textwidth}{@{}XXrl@{}}
\toprule
Lever & What it removes & Effect & Verdict \\
\midrule
master-cast reuse across micro-steps & 1,424 redundant cast\_f32\_bf16 launches a step & +0.18\% & kept \\
fused residual add + pre-norm & one [B,T,C] read per sublayer & -0.55\% & rejected \\
cuDNN dQ written direct in BF16 & rearrange\_n\_convert\_dq, 29.5 ms of a 4,008 ms step & -0.65\% & rejected \\
clip coefficient handed to the optimizer & 147 bmul\_f32 a step, 3.3 ms & -0.34\% & rejected \\
\bottomrule
\end{tabularx}
\end{table}

%% file: table_stage.tex
\begin{table}[t]
\caption{Where a training step's time goes, one card, 19 sequences per device.
numbat measured with \texttt{nsys}, PyTorch with \texttt{torch.profiler}, both at the
same batch and precision. Milliseconds per step.}
\label{tab:stage}
\centering\small
\begin{tabularx}{\textwidth}{@{}lrrX@{}}
\toprule
Stage & PyTorch & \numbat{} & What differs \\
\midrule
GEMMs (projections and head) & 206 & 202 & the same cuBLASLt/cutlass bf16 kernels \\
Attention & 26.3 & 30.8 & cuDNN fused flash against FlashAttention-2 \\
Cross-entropy forward and backward & 7.0 & 11.4 & two kernels against one Triton kernel \\
Embedding backward & 1.5 & 0.18 & one thread per element; PyTorch figure is the 1-2 ms range's midpoint \\
Residual adds, q/k/v copies, casts & fused & 11.0 & fused into neighbouring kernels by torch.compile; launched separately by numbat \\
Optimizer and clipping & 6.5 & 9.2 & per-tensor AdamW against fused multi-tensor \\
\midrule
Kernel total & 291 & 292 & \\
Wall clock & 287 & 296 & host-side launch and node overhead \\
\bottomrule
\end{tabularx}
\end{table}

%% file: table_clinical.tex
\begin{table}[t]
\caption{One epoch of clinical adaptation. Both legs ran the same recipe on the
same three cards, ONE AT A TIME and each alone on the box, the reference first so the
second could be gated against its curve. Data parallel over 3
ranks, one process per card: 48 examples an optimizer step.
Both score the same capped 512-example prefix of the
same held-out split, and the gate refuses the comparison when those counts differ, so
the two loss figures are one measurement rather than two similar ones. The step counts
differ by one because one epoch is 3{,}509.85 steps and the two legs round it
differently; both bests are at step 3{,}500, so the comparison is between the same
step. \textbf{No wall-clock row appears here on purpose}: the reference leg shared the
machine with an unrelated compile for part of its run and the other did not, so those
two numbers are not comparable. The speed comparison is \S\ref{sec:clinical}.}
\label{tab:clinical}
\centering\small
\begin{tabular}{@{}lrr@{}}
\toprule
 & \numbat{} & PyTorch + peft \\
\midrule
Steps & 3{,}509 & 3{,}510 \\
Evaluations & 14 & 15 \\
Held-out loss, first evaluation & 2.4251 & 2.4326 \\
\textbf{Held-out loss, best} & \textbf{2.1899} & 2.1941 \\
Held-out perplexity & 8.935 & 8.972 \\
Final training loss & 2.0466 & 2.1996 \\
\bottomrule
\end{tabular}
\end{table}